\documentclass{aa}  

\usepackage{float}
\usepackage{graphicx}
\usepackage{txfonts}
\usepackage{natbib}
\bibpunct{(}{)}{,}{a}{}{,}  
\usepackage{pdflscape}
\usepackage{orcidlink}
\usepackage{hyperref}
\hypersetup{breaklinks=true,
            colorlinks,
            filecolor=blue,
            linkcolor=blue,
            citecolor=blue,
            urlcolor=blue,
            anchorcolor=blue} 
\newcommand\feh{\ensuremath{[\mathrm{Fe}/\mathrm{H}]~}}
\newcommand\mh{\ensuremath{[\mathrm{M}/\mathrm{H}]~}}
\newcommand\afeh{\ensuremath{[\alpha/\mathrm{Fe}]~}}
\newcommand\dex{\ensuremath{~\mathrm{dex}}}

\defcitealias{massari2023}{Paper I}
\defcitealias{aguado-agelet25}{Paper II}
\defcitealias{ceccarelli2024b}{C24}

\begin{document}

   \title{Cluster Ages to Reconstruct the Milky Way Assembly (CARMA)}

   \subtitle{III. NGC 288 as the first Splashed globular cluster}
   
   \titlerunning{CARMA III.}

   \authorrunning{Ceccarelli, E., et al.}   
   
   \author{E. Ceccarelli
          \inst{1,2}
          \orcidlink{0009-0007-3793-9766},
          D. Massari
          \inst{1}
          \orcidlink{0000-0001-8892-4301},
          F. Aguado-Agelet
          \inst{3,4},         
          A. Mucciarelli
          \inst{2,1}
          \orcidlink{0000-0001-9158-8580},             
          S. Cassisi
          \inst{5,6},
          M. Monelli
          \inst{5,7},          
          E. Pancino
          \inst{8},          
          M. Salaris
          \inst{9}
          and
          S. Saracino
          \inst{8,9}        
          }         

   \institute{INAF - Osservatorio di Astrofisica e Scienza dello Spazio di Bologna, Via Gobetti 93/3, 40129 Bologna, Italy\\ \email{edoardo.ceccarelli3@unibo.it}
         \and
         Dipartimento di Fisica e Astronomia, Università degli Studi di Bologna, Via Gobetti 93/2, 40129 Bologna, Italy
         \and
         atlanTTic, Universidade de Vigo, Escola de Enxeñar\'ia de Telecomunicaci\'on, 36310, Vigo, Spain 
         \and
         Universidad de La Laguna, Avda. Astrof\'isico Fco. S\'anchez, E-38205 La Laguna, Tenerife, Spain     
         \and
         INAF - Osservatorio Astronomico di Abruzzo, Via M. Maggini, 64100 Teramo, Italy
         \and
         INFN – Sezione di Pisa, Università di Pisa, Largo Pontecorvo 3, 56127 Pisa, Italy
         \and
         INAF - Osservatorio Astronomico di Roma, Via Frascati 33, 00040 Monte Porzio Catone, Italy               
         \and
         INAF - Osservatorio Astrofisico di Arcetri, Largo E. Fermi 5, I-50125 Firenze, Italy
         \and
         Astrophysics Research Institute, Liverpool John Moores University, 146 Brownlow Hill, Liverpool L3 5RF, UK         
             }


 
  \abstract
  {The system of globular clusters (GCs) in the Milky Way (MW) comprises a mixture of both in situ and accreted clusters. Tracing the origin of GCs provides invaluable insights into the formation history of the MW. However, reconciling diverse strands of evidence is often challenging. A notable example is NGC 288, where despite significant efforts in the literature, the available chrono-chemodynamical data have yet to provide a definitive conclusion regarding its origin. On the one hand, all post-\textit{Gaia} dynamical studies indicate an accreted origin for NGC 288, pointing towards its formation taking place in the \textit{Gaia}-Sausage-Enceladus (GSE) dwarf galaxy. On the other hand, NGC 288 has been found to be 2.5 Gyr older than other GSE GCs at the same metallicity,  suggesting a different (and possibly in situ) origin. In this work, we address the unresolved question on the origin of NGC 288 by analysing its chrono-chemical properties in an unprecedentedly homogeneous framework. First, we compared the location of NGC 288 in the age-metallicity plane with that of other two GCs at similar metallicity, namely, NGC 6218 and NGC 6362, whose chemodynamical properties unambiguously identify them as in situ. The age estimates obtained within the homogeneous framework of the CARMA collaboration show that the three clusters are coeval,  reinforcing the contrast with the dynamical interpretation. Then, we derived the chemical composition of NGC 288 using UVES-FLAMES at VLT high-resolution spectroscopic archival data and compared the abundances with a sample of in situ and accreted clusters at similar metallicity. We found a consistency with the chemistry of in situ systems, especially in Si, Ti, Zn, and abundance ratios relative to Eu. To reconcile these results with its orbital properties, we propose a scenario where NGC 288 formed in the proto-disc of the MW and  was then dynamically heated by the interaction with the GSE merger. This is a fate that resembles that of proto-disc stars undergoing the so-called Splash event. Therefore, NGC 288  demonstrates the importance of a homogeneous chrono-chemodynamical information in the interpretation of the origin of MW GCs.}


   \maketitle
%

\section{Introduction}

The Milky Way (MW) has undergone a process of hierarchical accretion of smaller galaxies to build up its mass, as predicted by the $\Lambda$CDM cosmological paradigm \citep{white&frenk1991}. The \textit{Gaia} mission \citep{GC16,GC23} has enabled us to unveil the dynamical fossil record of those events that were hidden in the Galactic halo \citep{helmi2020}, such as the last major merger experienced by the MW, which occurred roughly 10 Gyr ago with the dwarf galaxy \textit{Gaia}-Sausage-Enceladus \citep[GSE, ][]{belokurov2018,helmi2018}. As a result of this impactful merger, the pre-existing MW disc became kinematically hotter, with a large fraction of its stars having had their orbits altered towards higher eccentricity. This process is sometimes referred to as the Splash \citep{dimatteo19,gallart19,belokurov2020}. 

It has been well-established that the most massive among the ingested dwarf galaxies harbored their own system of globular clusters (GCs) that were accreted alongside their stars \citep{penarrubia2009,kruijssen2019,bellazzini2020,malhan2022}. Thus, the study and characterization of GCs stand out as a powerful probe of the MW assembly processes. Dynamical information derived from the measurements of the \textit{Gaia} mission has allowed us to reconstruct the origin of each individual MW GC \citep{massari19,forbes2020,callingham2022,chen&gnedin2024}. However, this is not sufficient to achieve good accuracy in the associations, as coherent dynamical substructures are neither pure nor complete \citep{pagnini2023,chen&gnedin2024,mori24}. Clues about the origin of GCs provided by their orbital properties can be significantly enriched by incorporating the information on the age. As extensively shown in the literature, GCs that formed in accreted dwarf galaxies exhibit distinct age-metallicity relations (AMRs), which distinguish them from the trajectory followed by GCs formed in situ \citep{marin-franch09,forbes&bridges2010,dotter2010,leaman2013, massari19}, at least in the metal-intermediate and metal-rich regimes $(\mh > -1.3 \dex)$.

In this context, one peculiar case is that related to the GC NGC 288. In fact, the orbital properties of this cluster, which is located in the halo at a distance of 9.0 kpc from us \citep{vasiliev&baumgardt2021}, suggest an accreted origin from the GSE merger event \citep{massari19,forbes2020,callingham2022,chen&gnedin2024}. However, contrasting evidence has been found in the literature regarding its age. For example, several works found NGC 288 to be about 2 Gyr older than NGC 362 \citep[i.e. a GSE GC with similar metallicity,][]{green&norris1990,sarajedini1990,bellazzini2001,gontcharov2021}, whereas other works have reported an age difference between the two GCs always smaller than 1 Gyr \citep{stetson1996,dotter2010,vandenberg2013}. These inconsistencies might be caused by systematic effects due to different photometric systems, methods, and adopted theoretical models, that can easily add up to $\sim 2$ Gyr \citep{massari19}. Further, several studies hint at differences in the chemical composition of NGC 288 compared to accreted GCs, especially regarding the $\alpha$- \citep{horta2020} and the light elements \citep{belokurov&krastov2024}, pointing towards a potential in situ formation. Also, \citet{monty2023_2} highlighted a poor agreement between NGC 288 and GSE field stars in the \textit{s-} and \textit{r-} process elements, showing that a chemical evolutionary model of a GSE-like galaxy fails to reproduce the abundances measured in NGC 288. However, also in these cases, the chemical evidence comes from spectra covering different spectral ranges analysed with different methods and from the adoption of different models, which might result in significant systematic errors.

In this work, we examine the chrono-chemodynamical profile of NGC 288, aiming to resolve the apparent contradictions arising from its peculiar properties. To do so, we compare age and chemical composition of NGC 288 with those of in situ MW GCs in the same metallicity range ($\mh \sim -1.0 \dex$). The novelty of this study is that it is conducted within a completely homogeneous framework. On the one hand, GC ages are estimated by means of the tools developed within the Cluster Ages to Reconstruct the Milky Way Assembly (CARMA) collaboration (\citealt{massari2023}, \citealt{aguado-agelet25}, hereafter Paper I and II, respectively). On the other hand, the chemical analysis follows the same prescriptions and uses optical spectra obtained with the same instrument and with the same quality as in \citet[][C24 hereafter]{ceccarelli2024b}. This kind of analysis erases any potential offset introduced by the use of different approaches, leading to an interpretation of the observational evidence with the utmost precision.
   
\section{Comparison between the age of NGC 288 and in situ globular clusters}\label{sec:age_288}
\begin{table*}[!th]
 \caption{Results of the isochrone fitting for NGC 6218, NGC 6362 (from this work) and NGC 288 (from \citetalias{aguado-agelet25}).}\label{tab:results} 
 \centering
  \begin{tabular}{ccccc}
  \hline
  Name & [M/H] & $E(B-V)$ & $(m - M)_{0}$ & Age  \\
  & (dex) & (mag) & (mag) & (Gyr)  \\
  \hline \\
  \vspace{0.2cm}
  NGC 6218 & -1.17 \(^{+0.05}_{-0.02}\) & 0.20 \(^{+0.01}_{-0.01}\) & 13.57 \(^{+0.01}_{-0.02}\) & 13.42 \(^{+0.38}_{-0.40}\) \\ 
  \vspace{0.2cm}
  NGC 6362 & -0.92 \(^{+0.02}_{-0.02}\) & 0.07 \(^{+0.01}_{-0.01}\) & 14.43 \(^{+0.01}_{-0.01}\) & 13.86 \(^{+0.33}_{-0.36}\) \\   
  \hline \\
  \vspace{0.2cm}
  NGC 288 & -1.12 \(^{+0.08}_{-0.08}\) & 0.02 \(^{+0.01}_{-0.01}\) & 14.77 \(^{+0.01}_{-0.01}\) & 13.75 \(^{+0.28}_{-0.22}\) \\ 
  \hline  
 \end{tabular}
\tablefoot{The CMD fits and corner plots are shown in Figs. \ref{fig:ages_6218} and \ref{fig:ages_6362}. All results obtained from the CARMA project can be found at: \url{https://www.oas.inaf.it/en/research/m2-en/carma-en/}}
\end{table*}
   \begin{figure*}[!th]
   \centering
   \begin{minipage}{0.45\textwidth}
        \centering
        \includegraphics[width=1.0\textwidth]{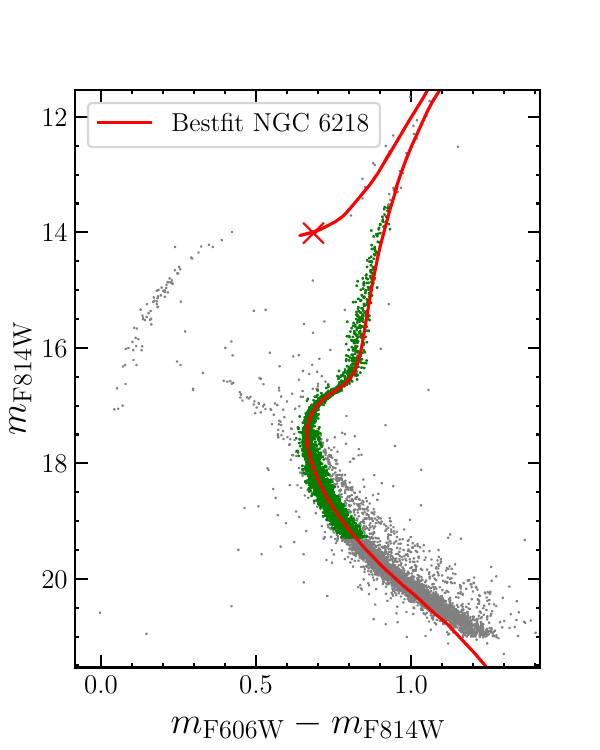}
        \includegraphics[width=1.0\textwidth]{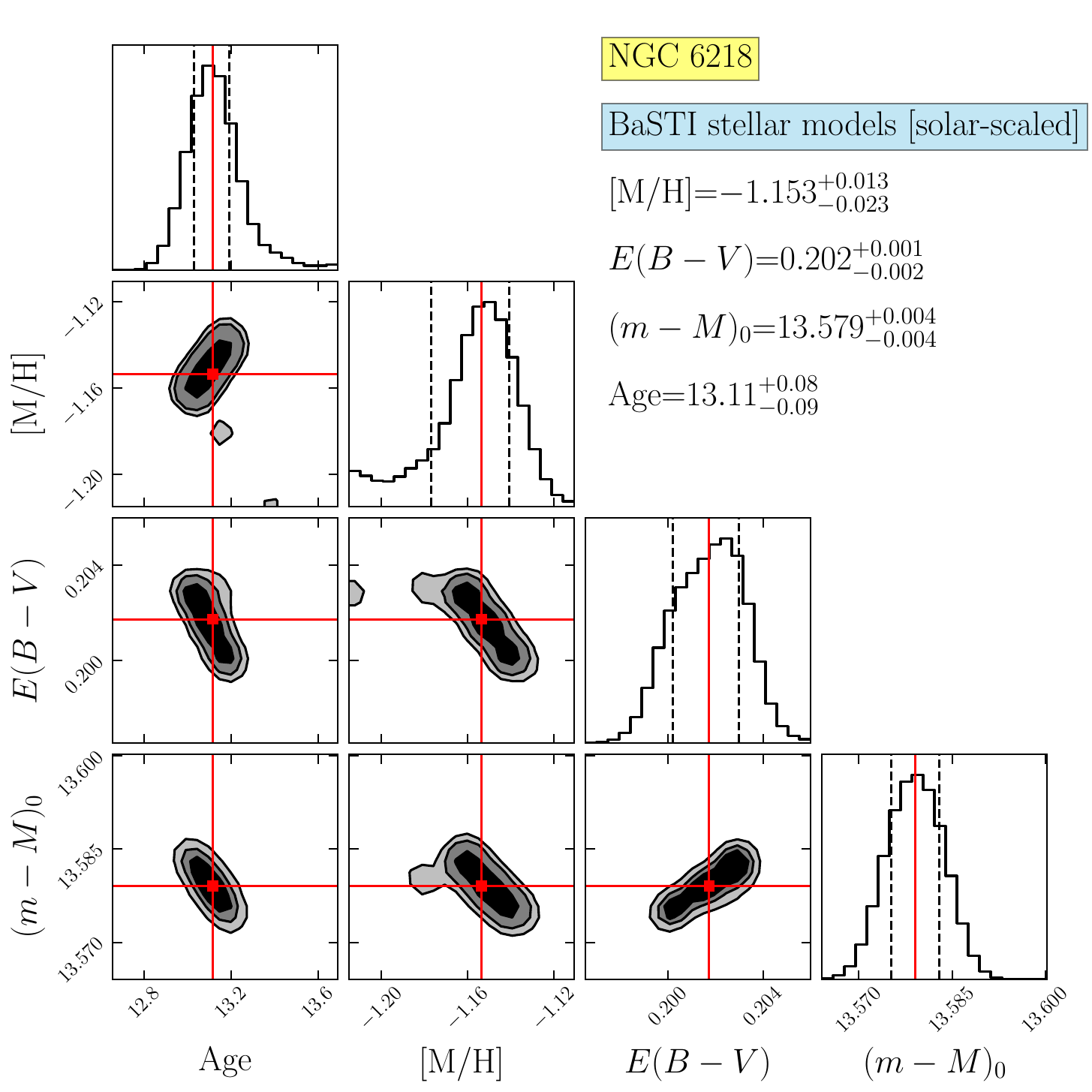}
   \end{minipage}\hfill
   \begin{minipage}{0.45\textwidth}
        \centering
        \includegraphics[width=1.0\textwidth]{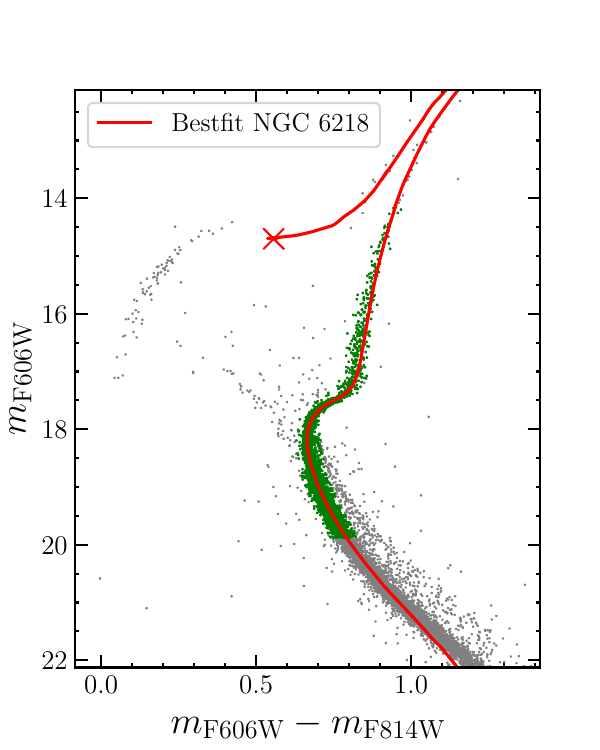}
        \includegraphics[width=1.0\textwidth]{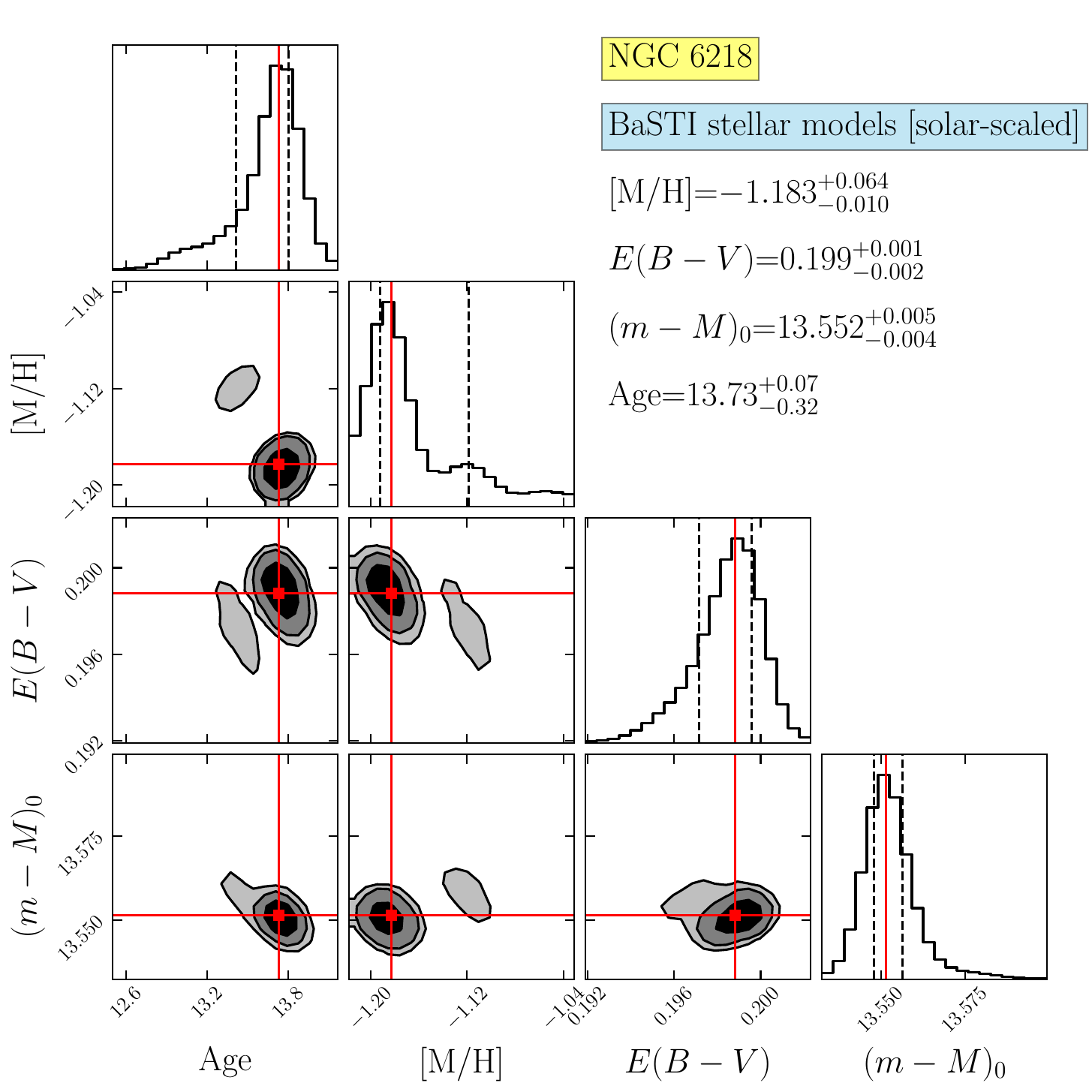}
   \end{minipage}
   \caption{Results for NGC 6218. Top-left: Best-fit model in the ($\it{m_{\mathrm{F814W}}}$, $\it{m_{\mathrm{F606W}}} - \it{m_{\mathrm{F814W}}}$) CMD. Top-right: Best-fit model in the ($\it{m_{\mathrm{F606W}}}$, $\it{m_{\mathrm{F606W}}} - \it{m_{\mathrm{F814W}}}$) CMD. Bottom-left: Posterior distributions for the output parameters and the best-fit solution, quoted in the labels, in the ($\it{m_{\mathrm{F814W}}}$, $\it{m_{\mathrm{F606W}}} - \it{m_{\mathrm{F814W}}}$) CMD. Bottom-right: Posterior distributions for the output parameters and the best-fit solution, quoted in the labels, in the ($\it{m_{\mathrm{F606W}}}$, $\it{m_{\mathrm{F606W}}} - \it{m_{\mathrm{F814W}}}$) CMD.}
              \label{fig:ages_6218}%
    \end{figure*}    
   \begin{figure*}[!th]
   \centering
   \begin{minipage}{0.45\textwidth}
        \centering
        \includegraphics[width=1.0\textwidth]{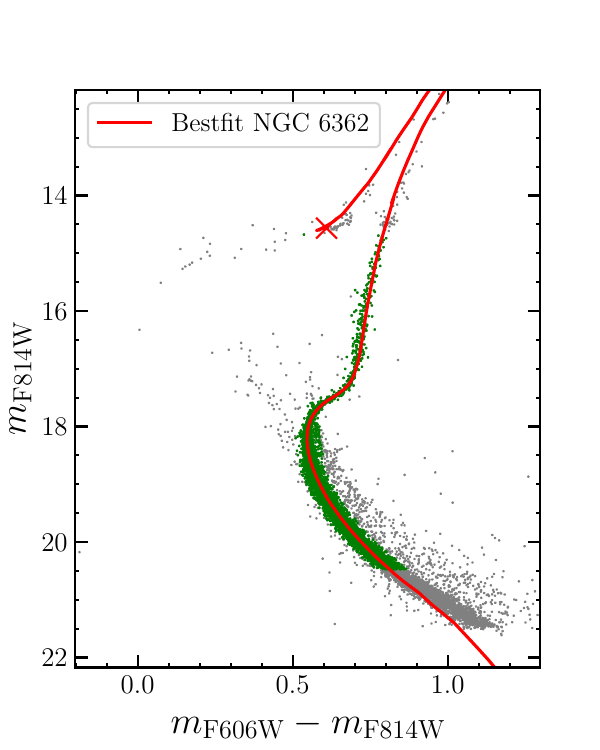}
        \includegraphics[width=1.0\textwidth]{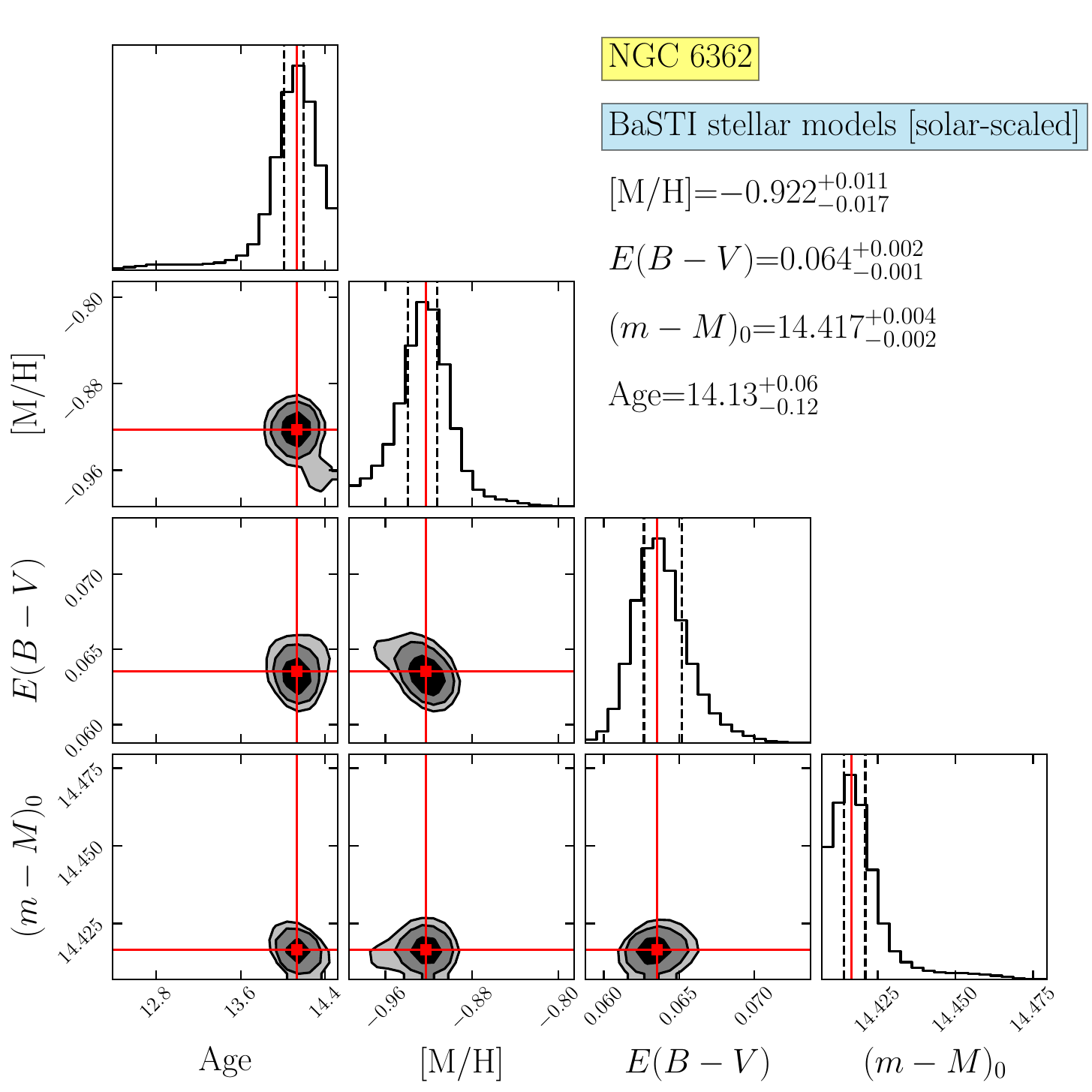}
   \end{minipage}\hfill
   \begin{minipage}{0.45\textwidth}
        \centering
        \includegraphics[width=1.0\textwidth]{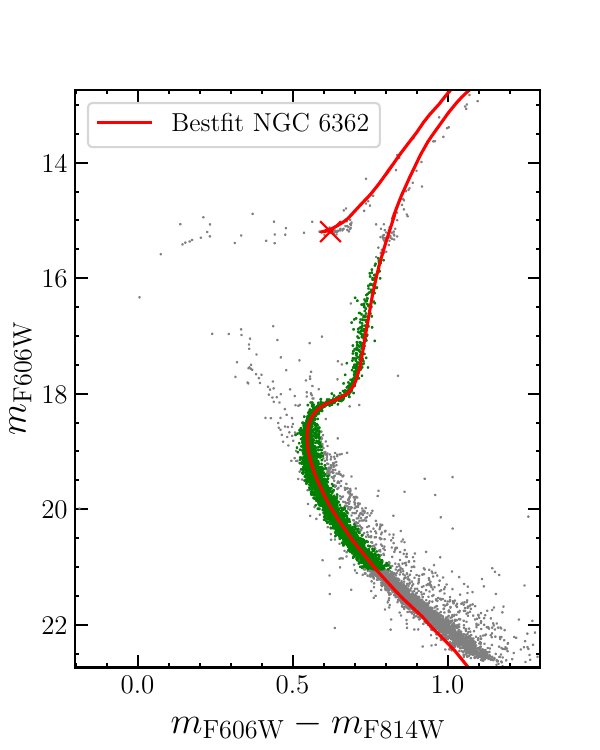}
        \includegraphics[width=1.0\textwidth]{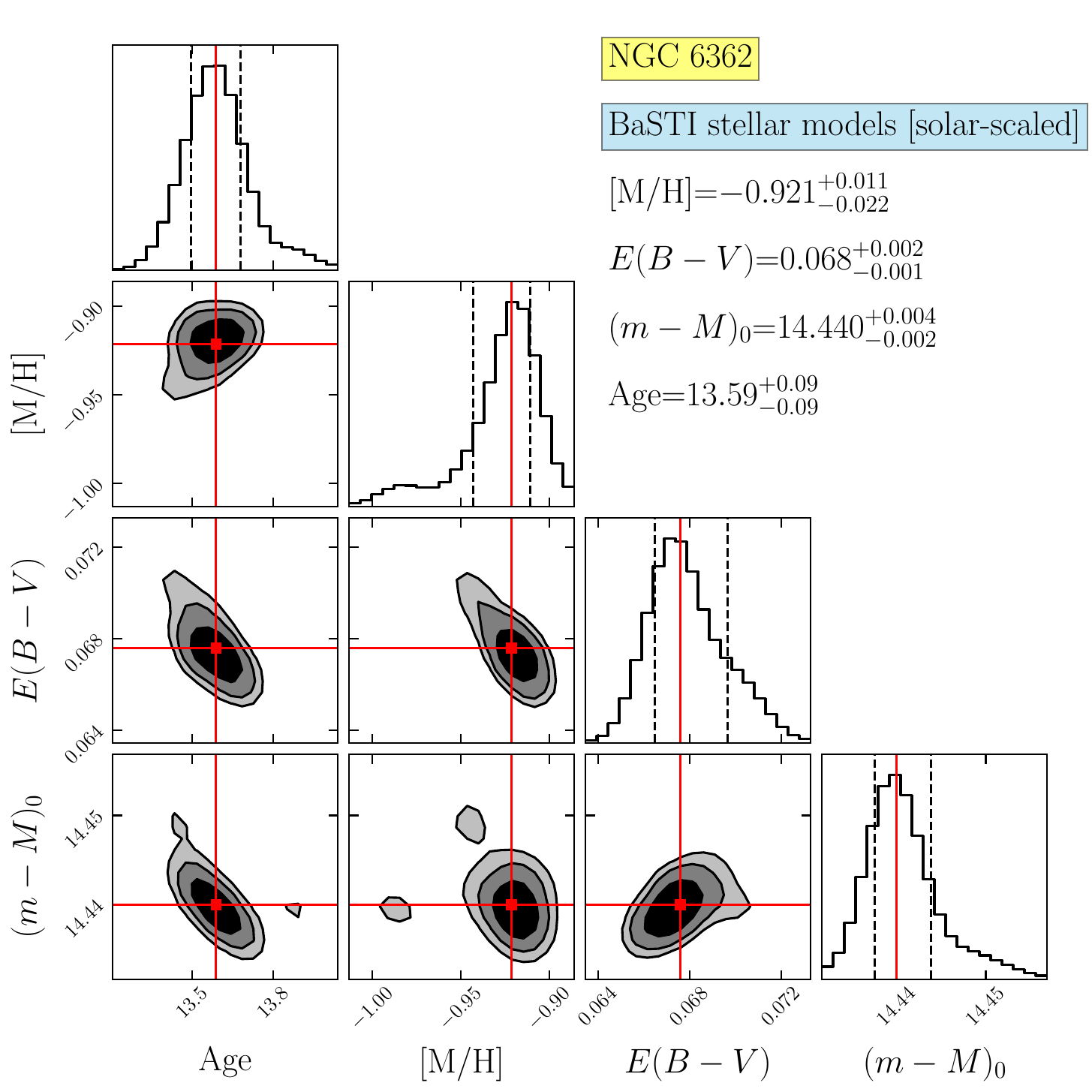}
   \end{minipage}
   \caption{Same as Fig. \ref{fig:ages_6218}, but for NGC 6362.}
              \label{fig:ages_6362}%
    \end{figure*}    
   \begin{figure}[!th]
   \centering
   \includegraphics[width=.45\textwidth]{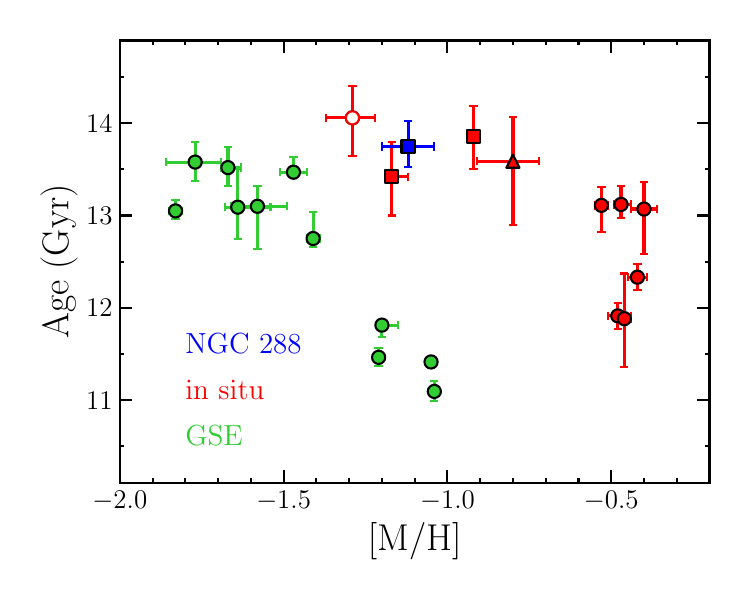} 
   \caption{Age-metallicity plane for all the GCs studied so far in the CARMA project. We plot the results for NGC 6218 and 6362 from this work as red filled squares. The position of NGC 288 from \citetalias{aguado-agelet25} is highlighted as a blue filled square. Red and green filled circles represent in situ and GSE GCs from \citetalias{massari2023} and \citetalias{aguado-agelet25}, respectively, while the red empty circle is NGC 6205. The red filled triangle is ESO452-11 from \citet{massari25}.}
              \label{fig:AMR}%
    \end{figure} 

The first objective of this work is to compare the age of NGC 288 as found in \citetalias{aguado-agelet25} with that of GCs at similar metallicity whose origin is clearly in situ. This type of relative comparison is, in fact, the best way to assess whether the hypothesis that NGC 288 is too old compared to GSE GCs to have been accreted is correct. According to the classifications by \citet{massari19} and \citet{callingham2022}, the only in situ disc GCs at $\mh \sim -1.0 \dex$ for which public Hubble Space Telescope (HST) photometry is available are NGC 6218 and NGC 6362. For both GCs, an in situ origin is supported by chemistry as well as by dynamics \citep[][\citetalias{ceccarelli2024b}]{belokurov&krastov2024}. 

To estimate the age of NGC 6218 and NGC 6362, we used photometry in the HST F606W and F814W filters from the HUGS survey \citep{piotto2015,nardiello2018}. In accordance with the typical prescriptions adopted by CARMA, we selected stars with a proper motion-based probability membership of $>$ 90\% and we applied a differential reddening correction \citep[][see Appendix \ref{app:DR} for details]{milone2012}. 

The isochrone fitting method employed to derive the age of NGC 6218 and NGC 6362 is presented in \citetalias{massari2023} (see Appendix \ref{app:if} for a brief summary of the methodology). The results of the isochrone fitting in both optical CMDs for these GCs are shown in Figs. \ref{fig:ages_6218} and \ref{fig:ages_6362} and summarised  in Table \ref{tab:results}. In the top panels of each figure, we show the best-fit isochrone superimposed on the two CMDs. Grey points represent the stars selected according to the methods described above (see also Appendix \ref{app:A}), while green points indicate stars specifically used for the fit. The posterior distributions of the parameters of the models are presented in bottom panels alongside the corner plot with the best fit values for [M/H], color excess, distance modulus, and age. To compare the global metallicity with literature results, we transformed [M/H] into the iron abundance following the prescriptions described in \citetalias{aguado-agelet25}. Thus, the photometric solution derived in this work translates to $\feh = -1.38 \dex$ for NGC 6218 and $\feh = -1.08 \dex$ for NGC 6362, which is consistent within $0.1 \dex$ with spectroscopic measurements \citep[][]{carretta09,massari17}. The median color excess values are $E(B-V) = 0.20 \pm 0.01$ and $E(B-V) = 0.07 \pm 0.01$ for NGC 6218 and NGC 6362, respectively. This  is also in agreement with previous literature estimates reported in \citet{harris2010}. Also, the distance modulus is well recovered, with differences with values listed in the \citet{harris2010} catalogue up to 0.15 mag, but fully consistent with results from \citet{baumgardt&vasiliev2021}.

In Fig. \ref{fig:AMR}, we present the AMR for the two GCs under study in this work (red filled squares) together with the results obtained for the metal-rich in situ GCs (red filled circles) in \citetalias{massari2023}, the in situ cluster ESO452-11 (red filled triangle) from \citet{massari25}, and the dynamically selected GSE GCs (green filled circles) in \citetalias{aguado-agelet25}. As  made evident by the figure, the age of NGC 288 is fully consistent within the uncertainties with that of the two in situ GCs analysed here. The mean age of the three GCs is 13.68 Gyr, with a very small dispersion of 0.19 Gyr. This in turn means that the three GCs are $\sim 2.5$ Gyr older than the sample of GSE GCs at the same metallicity, as expected for clusters that formed in an environment characterised by a higher star formation efficiency, compared to a dwarf galaxy such as GSE \citep{kruijssen2019,souza24,gonzalez-koda2025}. All the collected evidence therefore supports the interpretation of NGC 288 as born in situ in the MW, in contrast to its orbital properties \citep{massari19,callingham2022}. Finally, we note that one additional GSE system (i.e. NGC 6205, red empty circle) follows the same AMR as the three GCs described above, reinforcing the idea presented in \citetalias{aguado-agelet25}, namely, that it is a contaminant in the fully dynamical selected GSE sample with a likely in situ origin. As this GC is more metal-poor than those analysed here, a comprehensive analysis of NGC 6205 is left to a future dedicated work. 

\section{The chemical composition of NGC 288}\label{sec:chemical_composition}
   \begin{figure*}
   \centering
   \includegraphics[width=.59\textwidth]{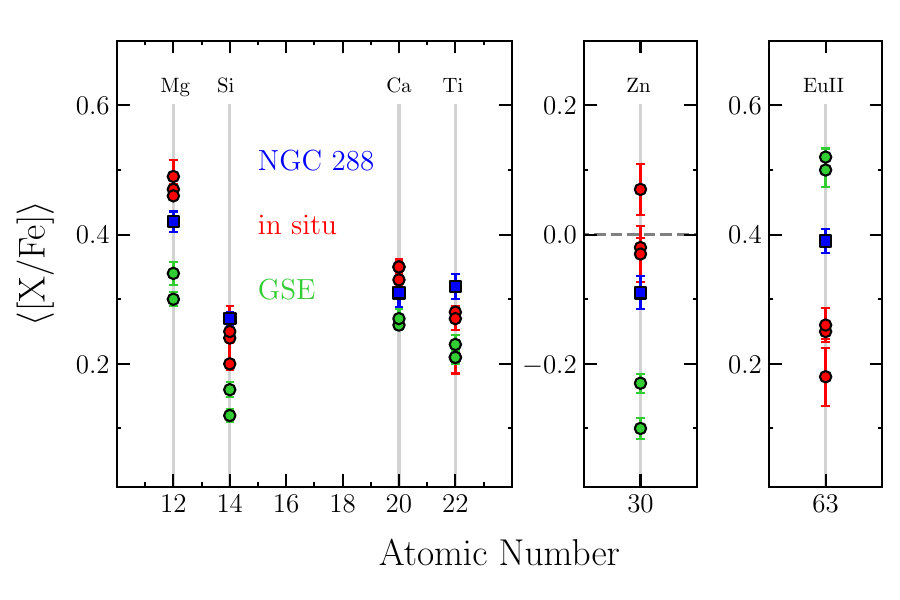} 
   \includegraphics[width=.395\textwidth]{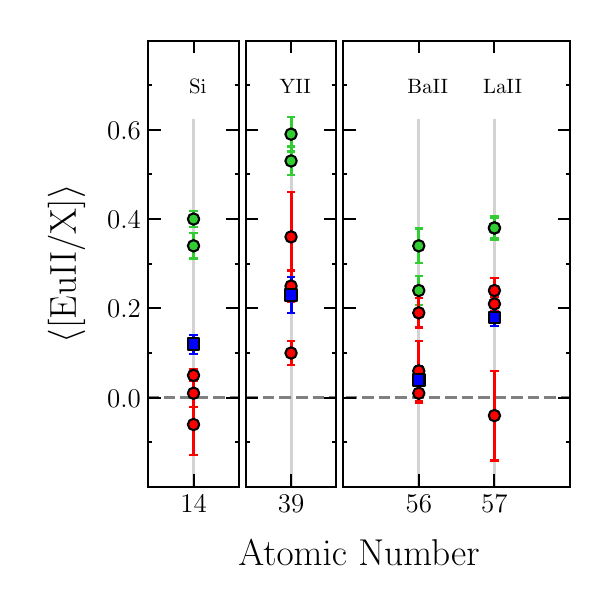}    
   \caption{Left panel: Difference of mean abundance ratios of the $\alpha$-elements [Mg/Fe], [Si/Fe], [Ca/Fe], [Ti/Fe], [Zn/Fe], and [EuII/Fe] between NGC 288 and GCs analyzed in \citetalias{ceccarelli2024b}. GCs are color coded according to their progenitors as in Fig. \ref{fig:AMR}. Right panel: Comparison of mean abundance ratios of EuII relative to other chemical elements (i.e. Si, YII, BaII, and LaII). Standard errors associated with the mean abundances are also reported.}
              \label{fig:chemistry}%
    \end{figure*}      
\begin{table}
  \caption{Average chemical abundance ratios for NGC 288.}\label{tab:abu_median}
  \centering
  \begin{tabular}{cc} 
   \hline             
Element & NGC 288   \\ 
 & (dex)  \\ 
\hline 
$\langle$[Fe/H]$\rangle$    &  -1.21    $\pm$   0.01 (0.03)     \\
$\langle$[Mg/Fe]$\rangle$   &   0.42    $\pm$   0.02 (0.05)     \\
$\langle$[Si/Fe]$\rangle$   &   0.27    $\pm$   0.01 (0.03)     \\
$\langle$[Ca/Fe]$\rangle$   &   0.31    $\pm$   0.02 (0.07)     \\ 
$\langle$[Ti/Fe]$\rangle$  &   0.32     $\pm$   0.02 (0.06)     \\
$\langle$[Zn/Fe]$\rangle$   &  -0.09    $\pm$   0.02 (0.08)     \\
$\langle$[YII/Fe]$\rangle$  &   0.16    $\pm$   0.04 (0.11)     \\
$\langle$[BaII/Fe]$\rangle$ &   0.35    $\pm$   0.03 (0.08)     \\
$\langle$[LaII/Fe]$\rangle$ &   0.21    $\pm$   0.01 (0.02)     \\
$\langle$[EuII/Fe]$\rangle$ &   0.39    $\pm$   0.02 (0.06)     \\
   \hline
  \end{tabular}
\tablefoot{The standard deviation is reported in parenthesis.}    
\end{table}

The combination of the chrono-dynamical information leads to contrasting results about the origin of NGC 288. In this context, the chemical composition of this system can plays a crucial role, since the abundances of stars reflect the chemical enrichment history of the environment where they formed. Thus, a detailed study of the chemical makeup of this GC can favour one of the two scenarios described above, offering clues to finally solving the puzzle regarding its origins. To do so, we followed the same approach used for the chronological information: a relative, systematic-free comparison between the abundances of NGC 288 and that of in situ and accreted GCs \citepalias{ceccarelli2024b}.

To perform the chemical analysis, we collected archival data of ten red giant branch (RGB) stars in NGC 288 that have been observed under the ESO-VLT Programme 073.D-0211 (PI: E. Carretta) using the multi-object spectrograph UVES-FLAMES \citep{pasquini02} mounted at the Very Large Telescope. The spectra were acquired using the Red Arm 580 CD3 grating with a spectral coverage between 4800 and 6800 \r{A}, along with a spectral resolution of R $\sim 40000$. The typical signal-to-noise ratio is S/N $\ge 55$ at 6000 \r{A}. The spectra were reduced using the dedicated ESO pipeline\footnote{\url{https://www.eso.org/sci/software/pipelines/}}. The sky background emission was  measured by observing empty sky regions and subsequently subtracted from each single stellar spectrum. To ensure homogeneity in the spectroscopic analysis, we followed the same approach used in \citetalias{ceccarelli2024b} to derive the atmospheric parameters and elemental abundances of target stars. For details of the chemical analysis, we refer to \citetalias{ceccarelli2024b} and references therein. A concise summary can be found in Appendix \ref{app:B}. The mean abundances for NGC 288 for key chemical elements that have proven their efficiency in disentangling accreted and in situ formation (i.e. $\alpha$-elements, Zn, Eu, \citetalias{ceccarelli2024b}) are listed in Table \ref{tab:abu_median}. The abundances of single stars can be found in Tables \ref{tab:alpha} and \ref{tab:n_capture}. The average iron content derived from neutral iron lines is $\feh = -1.21 \pm 0.01 \dex$, in good agreement with either spectroscopic \citep{carretta09,horta2020} and photometric solutions (see \citetalias{aguado-agelet25}).

In Fig. \ref{fig:chemistry}, we compare the average abundance of NGC 288 (blue symbol) and several MW GCs from \citetalias{ceccarelli2024b}, distinguishing between in situ (NGC 6218, NGC 6522 and NGC 6626, red symbols) and accreted from GSE (NGC 362 and NGC 1261, green symbols). As illustrated in the left panel, NGC 288 consistently shows enhancement ($\ge 0.1 \dex$) in $\alpha$-elements Si and Ti, compared to GSE GCs, with values comparable to those found in in situ GCs. A similar enhancement is also observed in Mg. Moreover, the average [Zn/Fe] ratio in NGC 288 is consistent (within the uncertainties) with that of in situ GCs. Given that absolute values of abundance ratios across different inhomogeneous studies may be subject to systematic zero-point offsets, our comparisons with literature are based on relative abundance differences derived within homogeneous analyses. The $\alpha$-element enhancements we find for NGC 288 relative to the GSE GC NGC 362 are consistent with results from previous studies \citep[][see also the discussion in the Appendix of \citetalias{ceccarelli2024b}]{shetrone2000,carretta09,carretta2010,carretta13,horta2020,monty23,monty2023_2}. Additionally, \citet{monty23,monty2023_2} also found a mild ($\sim 0.1 \dex$) [Zn/Fe] enhancement in NGC 288 relative to NGC 362. Furthermore, the differences detected in the $\alpha$- and neutron capture elements in NGC 288 and the in situ GC NGC 6218 are in agreement with those reported by \citet{carretta09,carretta2010,horta2020,schiappacasse-ulloa2024}.

Finally, we find that the [EuII/Fe] of NGC 288 falls in between the GSE and in situ groups, in agreement with literature results \citep{shetrone2000,monty24,schiappacasse2025}. However, since Eu is synthesized in various rare stellar events \citep[e.g., magneto-rotational supernovae, collapsars, neutron star mergers, see][]{cescutti2015,mosta2018,siegel2019}, its abundance is known to show significant intrinsic scatter, even among populations formed in the same environment \citep[see e.g.][]{venn2012,hill2019,matsuno2021,ou2024}. As shown in the right panel of Fig. \ref{fig:chemistry}, the chemical spaces where the consistency between NGC 288 and in situ GCs is more remarkable are those relative to Eu. It has been well-established that the [EuII/X] ratios are a powerful tool to discriminate between in situ and accreted GCs due to the different sites and timescales of production of this species compared to the $\alpha$-elements and to the light and heavy $s$-process elements \citep{kobayashi2020,molero2023,ou2024}. Indeed, chemical evolutionary models predict that a more massive galaxy, such as the MW compared to GSE, would be expected to have lower values of Eu compared to these chemical elements that are due  to either a higher star formation efficiency or the reduced impact of delayed $r$-process sources, such as neutron star mergers \citep{cote2019,palla25}. We find that NGC 288 is depleted in all [EuII/X] ratios compared to GSE GCs and always shows compatible values with in situ GCs from \citetalias{ceccarelli2024b}. The relative differences between these two groups are consistent with those observed for either GCs and field stars \citep{fishlock2017,aguado21,monty24,ou2024,schiappacasse2025}. This comparison demonstrates that the conditions of the gas where these clusters formed were extremely similar and coherent with the expectations for a galaxy like the MW in its earliest phases.  

\section{Discussion and conclusion}\label{sec:conclusion}

   \begin{figure}[!th]
   \centering
   \includegraphics[width=.45\textwidth]{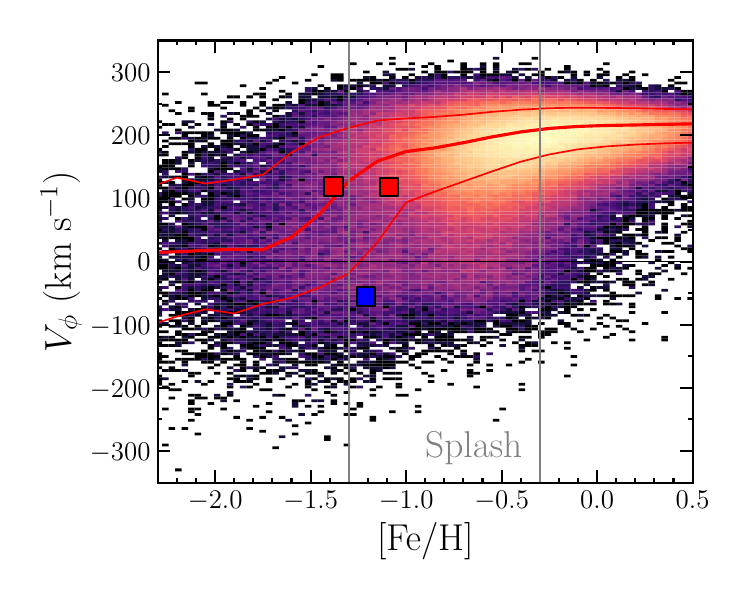} 
   \caption{Position of the NGC 288 (blue filled square) and the two in situ GCs (red filled squares) in the $V_{\phi} - \feh$ plane. In background we show the distribution of likely in situ stars from \citet{bellazzini2024}. The central red line highlights the median of the $V_{\phi}$ distribution, while the upper and lower lines trace the 16th and 84th percentiles, respectively. The grey vertical lines represent the upper and lower limits of the metallicity distribution function of Splash stars.}
              \label{fig:vphi_feh}%
    \end{figure} 

The picture that arises after combining all of these properties clearly suggests an in situ formation for NGC 288, which is at odds with its dynamical properties. Yet, just because these contradictory features have been determined in an extremely precise way, we can put forward a scenario that can accommodate all the observational evidences in a coherent picture.

In fact, we suggest that NGC 288 originally formed in situ within the MW, likely on a near-circular orbit with a low vertical height \citep{pfeffer2020}, until the merger with GSE took place. It is well established that this interaction likely dynamically heated up the pre-existing MW disc, driving stars onto more eccentric orbits \citep{helmi2018,dimatteo19}, a process commonly depicted as the Splash event \citep{belokurov2020}. In Fig. \ref{fig:vphi_feh}, we compare the position of NGC 288 (blue filled square) with that of NGC 6218 and NGC 6362 (red filled squares) in the $V_{\phi} - \feh$diagram, together with that of in situ stars\footnote{We note that NGC 288, NGC 6218, and NGC 6362 meet the selection criteria defined by these authors.} selected by \citet{bellazzini2024}. As depicted in this figure, NGC 6218 and NGC 6362 follow the distribution of in situ MW stars and GCs \citep{belokurov&krastov2024}, likely forming during the rapid spin-up of the MW disc \citep{belokurov22}. On the contrary, NGC 288 display a $V_{\phi}$ value that is consistent with that of stars in the metal-poor tail of the Splash at $\feh \sim -1.2$ dex. Our interpretation is further reinforced by the chemical properties of Splash stars, whose abundances of the $\alpha$-element are enhanced at about \afeh $= 0.3 \dex$ \citep{belokurov2020}, consistent with values measured for NGC 288. The fact that not all in situ GCs experienced similar dynamical perturbations from the GSE merger suggests that their initial spatial distribution within the Galaxy played a role in determining their response to this event. For example, the current larger apocentre in the orbit of NGC 288 suggests the possibility that this cluster could have already been located at a greater Galactocentric radius, compared to NGC 6218 and NGC 6362, making it more susceptible to the dynamical perturbations induced by the GSE merger. The idea that a major merger can similarly perturb the orbit of a GC has been supported also by cosmological simulations, which show that such events partially modify the kinematic distributions of in situ GCs \citep{keller20,chen&gnedin2022}. In particular, mergers can displace a non-negligible fraction of these clusters from the Galactic disc to the halo by increasing their orbital eccentricity and energy \citep{li&gnedin2019,trujillo-gomez2021}. Indeed, some old and metal-poor in situ GCs are expected to be found in the high energy and retrograde end of the $E-L_{\mathrm{z}}$ plane \citep{pfeffer2020}. Thus, this study may provide the first observational evidence of such a dynamical effect on an in situ MW GC, making NGC 288 the first example of a Splashed GC. In conclusion, these results mark NGC 288 as the first GC found to be dynamically consistent with an accreted progenitor, but yet not formed within it, demonstrating the effectiveness of combining homogeneous ages and chemical abundances and opening new avenues for understanding how GCs formed and evolved within our galaxy. 

\begin{acknowledgements}

    Based on observations collected at the ESO-VLT under the program 073.D-0211 (P.I. E. Carretta).

    This work has made use of data from the European Space Agency (ESA) mission \textit{Gaia} \url{https://www.cosmos.esa.int/gaia}), processed by the \textit{Gaia} Data Processing and Analysis Consortium (DPAC, \url{https://www.cosmos.esa.int/web/gaia/dpac/consortium}). Funding for the DPAC has been provided by national institutions, in particular the institutions participating in the \textit{Gaia} Multilateral Agreement. 

    EC is grateful to M. Bellazzini and A. Della Croce for useful discussions, comments and  support to this work.

    This research is funded by the project \textit{LEGO – Reconstructing the building blocks of the Galaxy by chemical tagging} (P.I. A. Mucciarelli), granted by the Italian MUR through contract PRIN 2022LLP8TK\_001.

    EC and DM acknowledge the support to this study by the PRIN INAF 2023 grant ObFu \textit{CHAM - Chemodynamics of the Accreted Halo of the Milky Way} (P.I. M. Bellazzini).

    DM acknowledges financial support from PRIN-MIUR-22: CHRONOS: adjusting the clock(s) to unveil the CHRONO-chemo-dynamical Structure of the Galaxy” (PI: S. Cassisi) granted by the European Union - Next Generation EU.

    DM acknowledges the support to activities related to the ESA/\textit{Gaia} mission by the Italian Space Agency (ASI) through contract 2018-24-HH.0 and its addendum 2018-24-HH.1-2022 to the National Institute for Astrophysics (INAF). 

    Co-funded by the European Union (ERC-2022-AdG, "StarDance: the non-canonical evolution of stars in clusters", Grant Agreement 101093572, PI: E. Pancino). Views and opinions expressed are however those of the author(s) only and do not necessarily reflect those of the European Union or the European Research Council. Neither the European Union nor the granting authority can be held responsible for them. 

\end{acknowledgements}

\bibliographystyle{aa}
\bibliography{aa54354-25.bib}

\begin{appendix}
\section{Age measurement}\label{app:A}

\subsection{Differential reddening correction}\label{app:DR}

\begin{figure}[!th]
   \centering
   \includegraphics[width=.48\columnwidth]{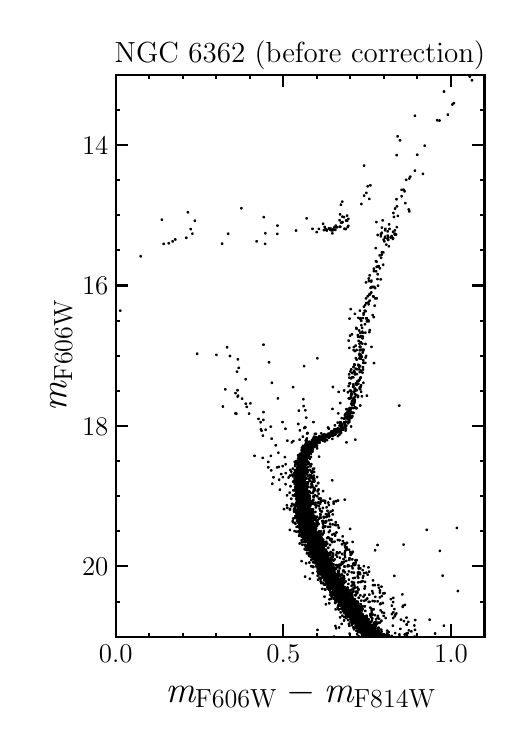} 
   \includegraphics[width=.48\columnwidth]{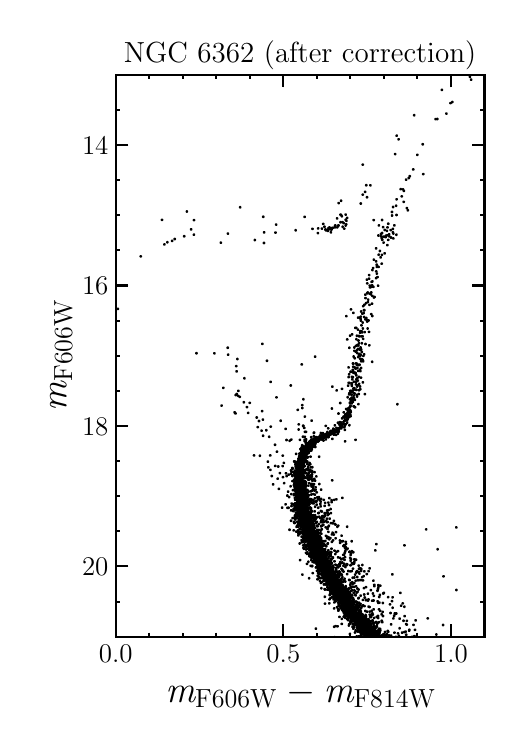}   
   \caption{Comparison of the CMDs of NGC 6362 before (left panel) and after (right panel) correcting for the effect of differential reddening.}
              \label{fig:DR_correction}%
    \end{figure}

\begin{figure}[!th]
   \centering
   \includegraphics[width=.45\textwidth]{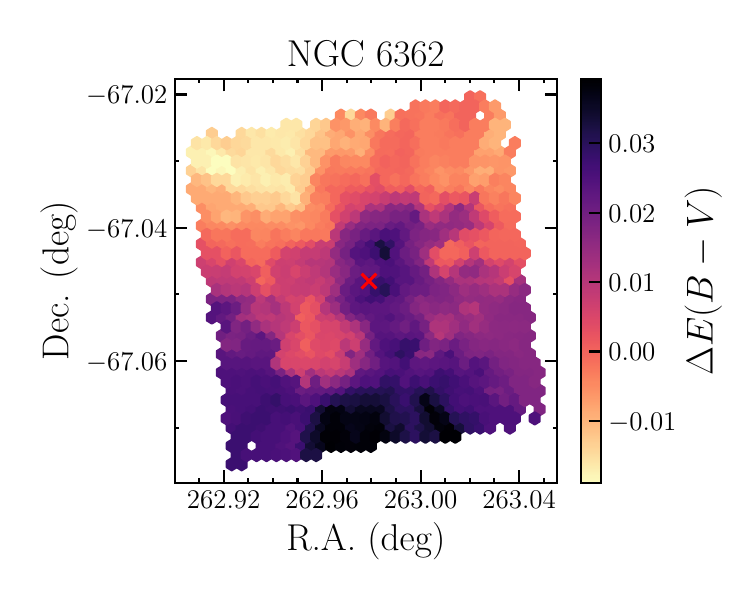} 
   \caption{Differential reddening map for the GC NGC 6362. The red cross indicates the centre of the cluster \citep{vasiliev&baumgardt2021}.}
              \label{fig:DR_map}%
    \end{figure}

  To ensure the highest quality data for the fitting process, we applied a differential reddening correction to the HST photometry of NGC 6218 and NGC 6362 following the approach described by \citet{milone2012}, after selecting for proper-motion based membership ($p>90\%$). We used RGB stars to define the reference sample, as their lower photometric errors make them preferable over main sequence stars, despite their smaller number. The differential reddening value ($\Delta E(B-V)$) associated with each star was derived as the median offset from the RGB mean ridge line, defined in a CMD oriented along the reddening vector, based on the 30 nearest reference stars. We employed the \citet{cardelli1989} extinction law, adopting $R_{\mathrm{V}} = 3.1$. This iterative process was repeated for each star until the residual $\Delta E(B-V)$ matched the typical photometric error, reaching convergence after two iterations. The effect of the differential reddening corrections on the CMD of NGC 6362 is illustrated in Fig. \ref{fig:DR_correction}, and the corresponding reddening map is displayed in Fig. \ref{fig:DR_map}.
\subsection{Isochrone fitting}\label{app:if}

  The age of NGC 6218 and NGC 6362 were estimated using the method described in \citetalias{massari2023}. We employed isochrones from the BaSTI library \citep{hidalgo18} spanning the age range from 6 to 15 Gyr and the [M/H] range from -2.5 to 0.0 dex (in steps of 0.1 Gyr and 0.01 dex, respectively). The CARMA isochrone fitting code was executed assuming Gaussian priors on the metallicity (solar-scaled models), color excess, and distance modulus centred on values provided in \citet{harris2010}, with dispersions of 0.1, 0.05 and 0.1, respectively. We recall that the use of solar-scaled models was deliberately chosen to avoid imposing assumptions about the $\alpha$-element abundance. On top of this, we chose to use photometric data in optical bands, where the equivalence between solar-scaled and $\alpha$-enhanced models at the same global metallicity has been robustly demonstrated \citep[e.g.][]{salaris93,cassisi04}. The code was ran on both the ($\it{m_{\mathrm{F814W}}}$, $\it{m_{\mathrm{F606W}}} - \it{m_{\mathrm{F814W}}}$) CMD and the ($\it{m_{\mathrm{F606W}}}$, $\it{m_{\mathrm{F606W}}} - \it{m_{\mathrm{F814W}}}$) CMD. In Table \ref{tab:results} we report the average value of the two runs and the associated asymmetric uncertainties calculated in order to enclose both the upper and lower limits of each run.


\section{Chemical analysis}\label{app:B}
\subsection{Stellar parameters}\label{app:sp}

%
\begin{table*}
\caption{Information on the targets.}          
\label{tab:SP}      
\centering          
\begin{tabular}{ccccc}  
\hline      
Star ID & ID Gaia DR3 & $T_{\mathrm{eff}}$ & log $g$ & $v_{\mathrm{t}}$  \\ 
 & & (K) & (dex) & (km $\mathrm{s^{-1}}$)  \\ 
\hline 
200011   &       2342906141734490880     &      4486     &      1.32     &       1.4       \\
200055   &       2342907825361664128     &      4844     &      2.00     &       1.5       \\
200050   &       2342903083717827200     &      4823     &      1.96     &       1.4       \\
200004   &       2342903118077555840     &      4242     &      0.87     &       1.5       \\
200002   &       2342903045061492224     &      4093     &      0.65     &       1.5       \\
200009   &       2342907580546992128     &      4458     &      1.24     &       1.5       \\
200038   &       2342903045061492224     &      4745     &      1.78     &       1.4       \\
200021   &       2342904698625661824     &      4616     &      1.53     &       1.4       \\
200061   &       2342908340757896704     &      4908     &      2.08     &       1.4       \\
200020   &       2342904423747772416     &      4597     &      1.49     &       1.4       \\
\hline                  
\end{tabular}
\tablefoot{We list the star ID from previous literature works \citep[e.g.][]{carretta09}, ID from \textit{Gaia} DR3, effective temperature, surface gravity, and microturbulent velocity. Typical uncertainties are of the order of $100$ K, 0.01 dex, and 0.2 km $\mathrm{s^{-1}}$ for $T_{\mathrm{eff}}$, log $g$, and $v_{\mathrm{t}}$ respectively.}
\end{table*}
%

   We derived the effective temperature ($T_{\mathrm{eff}}$) and the surface gravity (log $g$) using the \textit{Gaia} Data Release 3 \citep[][]{GC23} photometric dataset exploiting the $\mathrm{(BP-RP)_{0}}$ - $T_{\mathrm{eff}}$ relation provided by \citet{mucciarelli21}. We assumed a color excess $E(B-V)$ from the \citet{harris2010} catalogue and followed the iterative prescription described in \citet{GC18_extinction} to efficiently remove the effects of the extinction on the observed $\mathrm{(BP-RP)}$ color. Internal errors in $T_{\mathrm{eff}}$ stem from the propagation of the uncertainties in the photometry, reddening and assumed color - $T_{\mathrm{eff}}$ relation and they are of the order of 80 - 100 K. Surface gravities (log $g$) have been estimated from the Stefan-Boltzmann relation, assuming the photometric $T_{\mathrm{eff}}$ and a stellar mass of 0.8 $\mathrm{M_{\odot}}$, the G-band bolometric corrections provided by \citet{andrae18}, and a distance of $8.99 \pm 0.09$ kpc \citep{baumgardt&vasiliev2021}, consistent with the procedure described in \citetalias{ceccarelli2024b}. We note that the distance by \citet{baumgardt&vasiliev2021} is fully consistent with that derived in \citetalias{aguado-agelet25}. We assume a conservative estimate of the uncertainty on the surface gravity of $0.1 \dex$. In the end, we calculated microturbulent velocities ($v_{\mathrm{t}}$) by erasing any trend between the measured iron abundances and reduced equivalent widths. Uncertainties on $v_{\mathrm{t}}$ are typically lower than 0.2 km $\mathrm{s^{-1}}$. All the atmospheric parameters are listed in Table \ref{tab:SP}.

\subsection{Derivation of abundances}

\begin{table*}
  \caption{Iron and $\alpha$-elements abundances for target stars in NGC 288.}\label{tab:alpha}
  \centering
  \begin{tabular}{cccccc} 
   \hline             
 Star ID & [Fe/H] & [Mg/Fe] & [Si/Fe] & [Ca/Fe] & [Ti/Fe] \\ 
 & (dex) & (dex) & (dex) & (dex) & (dex) \\ 
\hline 
200011                   &      -1.19   $\pm$   0.09     &      0.49    $\pm$   0.04     &       0.29    $\pm$   0.10     &      0.34    $\pm$   0.05     &      0.33    $\pm$   0.09    \\ 
200055                   &      -1.15   $\pm$   0.11     &      0.38    $\pm$   0.03     &       0.25    $\pm$   0.10     &      0.19    $\pm$   0.04     &      0.24    $\pm$   0.05    \\ 
200050                   &      -1.21   $\pm$   0.10     &      0.44    $\pm$   0.02     &       0.28    $\pm$   0.09     &      0.40    $\pm$   0.06     &      0.28    $\pm$   0.06    \\ 
200004                   &      -1.26   $\pm$   0.07     &              -                    &   0.27    $\pm$   0.11     &      0.41    $\pm$   0.07     &       0.42    $\pm$   0.13    \\ 
200002                   &      -1.26   $\pm$   0.06     &              -                    &   0.32    $\pm$   0.12     &      0.40    $\pm$   0.10     &       0.43    $\pm$   0.15     \\ 
200009                   &      -1.22   $\pm$   0.10     &      0.45    $\pm$   0.04     &       0.22    $\pm$   0.12     &      0.29    $\pm$   0.05     &      0.33    $\pm$   0.09     \\ 
200038                   &      -1.17   $\pm$   0.11     &      0.40    $\pm$   0.04     &       0.26    $\pm$   0.10     &      0.27    $\pm$   0.04     &      0.28    $\pm$   0.06     \\ 
200021                   &      -1.21   $\pm$   0.10     &      0.43    $\pm$   0.03     &       0.25    $\pm$   0.10     &      0.30    $\pm$   0.05     &      0.32    $\pm$   0.07     \\ 
200061                   &      -1.22   $\pm$   0.10     &      0.33    $\pm$   0.03     &       0.28    $\pm$   0.08     &      0.23    $\pm$   0.05     &      0.26    $\pm$   0.05     \\ 
200020                   &      -1.21   $\pm$   0.10     &      0.43    $\pm$   0.03     &       0.27    $\pm$   0.10     &      0.26    $\pm$   0.04     &      0.29    $\pm$   0.08     \\ 
\hline 
  \end{tabular}
\end{table*}
\begin{table*}
  \caption{Same as Table \ref{tab:alpha}, but for Zn, YII, BaII, LaII, and EuII.}\label{tab:n_capture}
  \centering
  \begin{tabular}{ccccccc} 
   \hline             
 Star ID & [Zn/Fe] & [YII/Fe] & [BaII/Fe] & [LaII/Fe] & [EuII/Fe]  \\ 
 & (dex) & (dex) & (dex) & (dex) & (dex)   \\ 
\hline 
200011                   &      -0.09   $\pm$   0.16     &      0.38    $\pm$   0.15     &       0.49    $\pm$   0.17     &      0.20    $\pm$   0.05     &      0.43    $\pm$   0.05      \\
200055                   &      -0.03   $\pm$   0.14     &      -0.01   $\pm$   0.15     &       0.17    $\pm$   0.17     &      0.21    $\pm$   0.05     &      0.35    $\pm$   0.06      \\
200050                   &      -0.06   $\pm$   0.13     &      0.15    $\pm$   0.14     &       0.39    $\pm$   0.16     &      0.17    $\pm$   0.05     &      0.37    $\pm$   0.06      \\
200004                   &      -0.21   $\pm$   0.15     &      0.19    $\pm$   0.15     &       0.32    $\pm$   0.15     &      0.20    $\pm$   0.05     &      0.40    $\pm$   0.05      \\
200002                   &      -0.24   $\pm$   0.14     &      0.31    $\pm$   0.15     &       0.37    $\pm$   0.16     &      0.22    $\pm$   0.05     &      0.47    $\pm$   0.05      \\
200009                   &      -0.06   $\pm$   0.16     &      0.10    $\pm$   0.15     &       0.27    $\pm$   0.16     &      0.20    $\pm$   0.05     &      0.36    $\pm$   0.05      \\
200038                   &      0.02    $\pm$   0.14     &      0.04    $\pm$   0.13     &       0.31    $\pm$   0.15     &      0.22    $\pm$   0.05     &      0.40    $\pm$   0.06      \\
200021                   &      -0.08   $\pm$   0.15     &      0.10    $\pm$   0.15     &       0.35    $\pm$   0.16     &      0.21    $\pm$   0.04     &      0.46    $\pm$   0.05      \\
200061                   &      -0.01   $\pm$   0.14     &      0.15    $\pm$   0.14     &       0.39    $\pm$   0.15     &      0.26    $\pm$   0.06     &      0.28    $\pm$   0.06      \\
200020                   &      -0.13   $\pm$   0.15     &      0.22    $\pm$   0.15     &       0.39    $\pm$   0.16     &      0.18    $\pm$   0.05     &      0.34    $\pm$   0.05      \\
   \hline
  \end{tabular}
\end{table*}

   We employed \texttt{ATLAS9} \citep{kurucz} model atmospheres computed under the assumptions of plane-parallel geometry, hydrostatic and radiative equilibrium, and local thermodynamic equilibrium for all the chemical elements, starting from an $\alpha$-enhanced model. Chemical abundances of Mg, Si, Ca, Ti, and Zn were obtained comparing theoretical and observed equivalent widths (EWs), measured with the code \texttt{DAOSPEC} \citep{daospec} exploiting the tool \texttt{4DAO} \citep{4dao}, using the code \texttt{GALA} \citep{gala}. Atomic lines for the chemical elements that have saturated hyperfine/isotopic splitting transitions (YII, BaII, LaII, and EuII) were analysed through spectral synthesis using the proprietary code \texttt{SALVADOR}. 

   The final abundance ratios are scaled to the solar values provided by \citet{grevesse1998}, that is the solar composition assumed when computing \texttt{ATLAS9} model atmospheres \citep{castelli&kurucz2003}. We report chemical abundances for the 10 RGB stars analysed in this work in Tables \ref{tab:alpha} and \ref{tab:n_capture}.

   The errors on the abundance ratios have been estimated as the squared sum of two components (see \citetalias{ceccarelli2024b} for a complete discussion): (i) internal errors due to the EW measurement and (ii) errors arising from the computation of atmospheric parameters. Thus, the uncertainties have been estimated as
   
   \begin{flalign*}
      &\sigma_{\feh}  = 
      \sqrt{\frac{\sigma_{Fe}^{2}}{N_{Fe}}  +
      (\delta_{Fe}^{T_{\mathrm{eff}}})^{2}  +
      (\delta_{Fe}^{\mathrm{log} \; g})^{2} + 
      (\delta_{Fe}^{v_{\mathrm{t}}})^{2}}  ,\\ \nonumber
      & \\ \nonumber
      &\sigma_{\mathrm{[X/Fe]}} = \\ 
      & \sqrt{\frac{\sigma_{X}^{2}}{N_{X}} + \frac{\sigma_{Fe}^{2}}{N_{Fe}} + (\delta_{X}^{T_{\mathrm{eff}}} - \delta_{Fe}^{T_{\mathrm{eff}}})^{2} + (\delta_{X}^{\mathrm{log} \; g} - \delta_{Fe}^{\mathrm{log} \; g})^{2} +(\delta_{X}^{v_{\mathrm{t}}} - \delta_{Fe}^{v_{\mathrm{t}}})^{2}},
   \end{flalign*}
   where $\sigma_{X,Fe}$ is the dispersion around the mean of chemical abundances, $N_{X,Fe}$ is the number of used lines and $\delta_{X,Fe}^{i}$ are the abundance differences obtained by varying the parameter $i$.    


\end{appendix}
\end{document}